\documentclass[a4paper,11pt]{article}
\pdfoutput=1 % if your are submitting a pdflatex (i.e. if you have
\usepackage{jinstpub} % for details on the use of the package, please
\usepackage{amsmath}
\usepackage{graphicx}
\usepackage[english]{babel}
\usepackage[latin1]{inputenc}
\usepackage[T1]{fontenc}
\usepackage{adjustbox}
\usepackage{graphicx}
\usepackage{textcomp}

\title{Expected performance of the TT-PET scanner}

\author[a,b,1]{E.~Ripiccini, \note{Corresponding author.}}
\author[a,b]{D. Hayakawa, }
\author[a]{G. Iacobucci, }
\author[a,c]{M. Nessi, }
\author[c]{E. Nowak, }
\author[a]{L. Paolozzi, }
\author[b]{O. Ratib, }
\author[a]{P. Valerio  }
\author[a]{and D. Vitturini}

\affiliation[a]{University of Geneva, Rue du G\'en\'eral-Dufour 24, Geneva, Switzerland}
\affiliation[b]{Institute of Translational Molucular Imaging (ITMI), University of Geneva Geneva, Switzerland}
\affiliation[c]{CERN\unskip, Geneva, Switzerland}

\emailAdd{emanuele.ripiccini@unige.ch}

\abstract{The TT-PET collaboration is developing a small animal TOF-PET scanner based on silicon detectors featuring 30 ps RMS time resolution and intended to be inserted in an existing MRI scanner. The TT-PET scanner makes use of a stack of layers of high-Z photon-converter and 100 $\mathrm{\mu m}$ thick
silicon sensors, to achieve a scanner with 0.5 $\mathrm{\times}$ 0.5 $\mathrm{\times}$ 0.2 $\mathrm{mm^{3}}$ granularity, with precise depth-of-interaction measurement.
In this paper we present the results of the Monte Carlo studies for the expected data rate, time resolution on the TOF measurements, spatial resolution and image reconstruction with and without the use of the timing information. Most of the studies have been performed according to the international standards used to assess the performance of small-animal PET system.}

\begin{document}
\maketitle
\flushbottom

\section{Introduction}
\label{sec:intro}

The TT-PET scanner is a structure of stacked silicon sensors, providing very high radial resolution, as opposed to high density, thick scintillating crystals (e.g.: LYSO, LSO, or BGO) commonly used in commercial scanners. PET scanners that exploit the high granularity of silicon detectors to significantly improve the spatial resolution were developed \cite{sil1} \cite{sil2}. The sensor we are developing combines the high granularity to a fast time response. As explained in \cite{trento}, a high time resolution is achieved by using thin silicon layers. Due to the low density of the silicon, a high Z converter that absorbs the annihilation photons, coming from the object under study, and generating electrons detectable by the sensor. 
The $ \gamma $-ray detection efficiency depends linearly on the probability of the the photon to convert and the probability of the produced electron to reach the silicon sensor.
The configuration that maximises the detection efficiency was defined by a detailed Monte Carlo simulation.
The scanner is divided in 16 towers around the scanner axis, with cooling blocks installed between the towers.
This work is part of a multidisciplinary project aimed to develop a new generation of high-performance small-animal PET inserts adaptable to existing MRI scanners, available at the Cantonal Hospital of Geneva, to produce high-resolution hybrid PET-MRI images. The initial phase of the project focuses on a first design for animal imaging and is intended to be extended to human brain imaging applications with a larger unit.

\section{Scanner layout}
The scanner has a cylindrical structure with an internal radius of 1.8 cm, external radius of 4.2 cm and a total length of 5 cm. The ring is made of 16 wedge-shaped units called "towers", (see Fig.~\ref{entirescanner}). Each tower is formed by:

\begin{figure}[h!]
\centering
\includegraphics[width=0.6\linewidth]{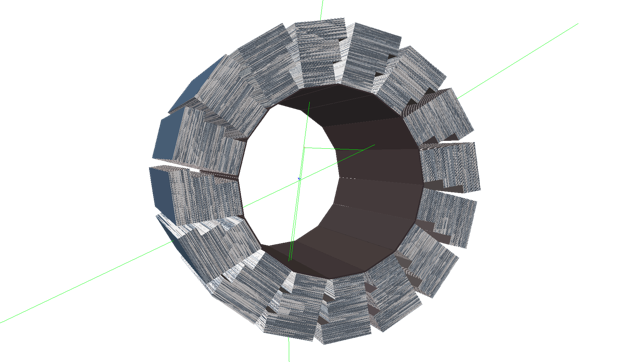}
\caption{Scanner geometry, showing the 16 towers made of 60 conversion and detection layers. Three detection layer widths are radially used.}
\label{entirescanner}
\end{figure}

\begin{figure}[h!]
\centering
\includegraphics[width=0.6\linewidth]{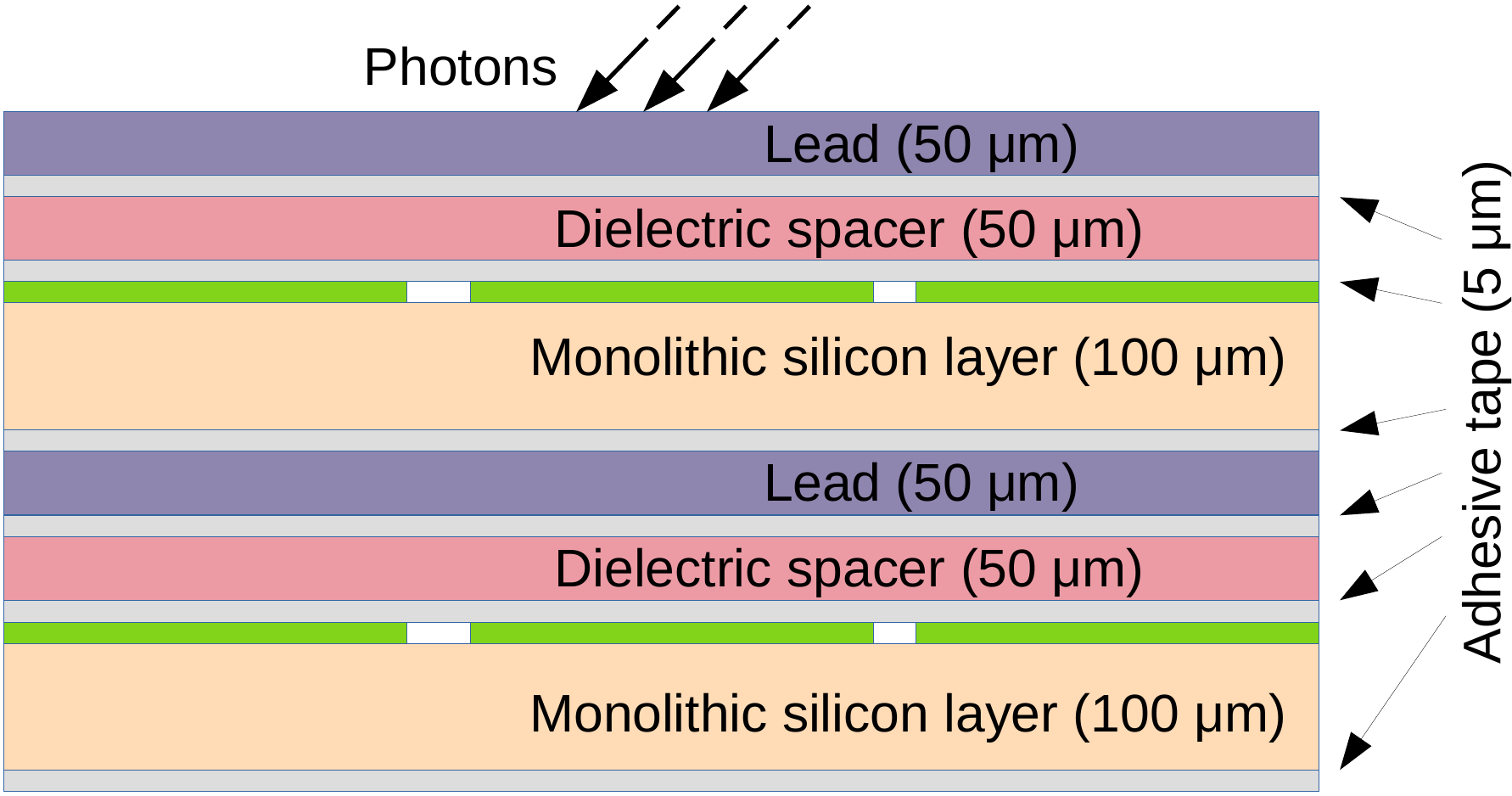}
\caption{Representation of two consecutive detection layers. }
\label{singlelayer}
\end{figure}
\begin{figure}[h!]
\centering
\includegraphics[width=0.7\linewidth]{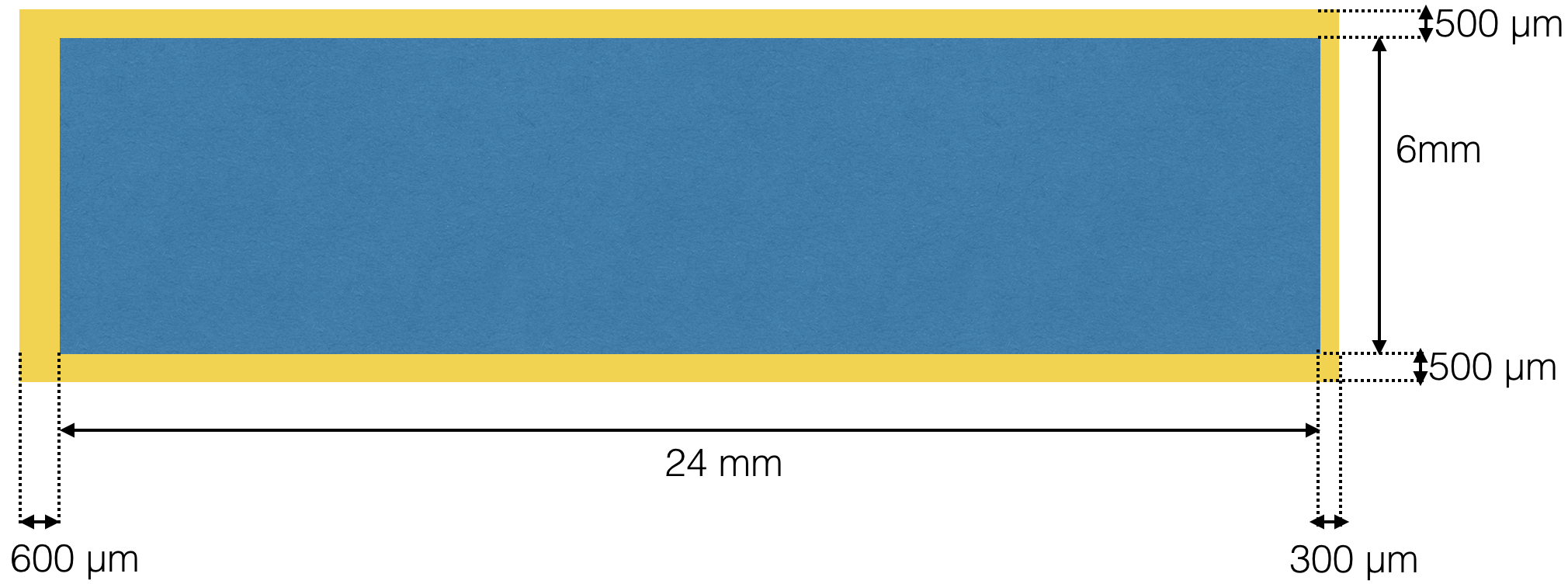}
\caption{The chip structure for the first super layer is reported. The blue region represents the sensitive area of the chip, while the yellow frame is the part dedicated to the front-end electronics.}
\label{chip}
\end{figure}

\begin{itemize}

\item 60 detection layers of three different sensor widths (7, 9, 11 mm).
\item Each detection layer is made of a 50 \textmu{}m thick converter layer glued to a 100 \textmu{}m thick silicon layer, as shown Fig.~\ref{singlelayer}.
\item Each silicon layer is formed by two 2.5 cm long chips like the one shown in Fig.~\ref{chip}. The active area of the chip is segmented in detection pixels of $ 0.5\times0.5 $ $ \mathrm{mm}^{2} $ .

\end{itemize}

The three chip types differ in the number of pixels, starting from the bottom the chip contains: 576, 768 and 960 pixels. The total number of read-out channel in the scanner is 1'474'560.

A dedicated Monte Carlo simulation showed that for 511 keV $\gamma$-rays produced in a point in the center of the scanner and isotropically emitted, the geometrical acceptance is 78$ \% $. 

The sensor is a novel particle detector based on silicon technology, featuring time measurement precision of 30 ps RMS for electrons at the 100 keV energy scale. 
Given the large amount of channels of the scanner, a monolithic full custom ASIC containing both the sensors and the front-end electronics has been designed \cite{peric} to simplify the signal extraction and power distribution. The signal from each pixel is routed to the chip periphery where the front-end electronics is located.  All the discriminated signals on one chip are finally multiplexed to a TDC (featuring 20 ps time binning). A detailed description of the front-end electronics can be found in \cite{asicpaper}.

A single tower is organised in 12 stacks, called super-modules, of 5 detection layers. Chips belonging to the same super-module are connected through wire bonding to a thin flex PCB that communicates with a custom electronic board, called "tower control", that provides temporary data storage, generates low and high voltage supplies and distributes synchronisation signal (160 MHz). 
A serial link running at 50 Mbps allows data to be readout from the super-module. Chips can be run in data-driven mode, where each hit is immediately sent out to the tower control module, or in triggered mode, where hits are stored in a local buffer and are sent out only after a request. For each hit, a pixel and chip address is read out in order to identify the hit position, in addition to timing data for both time-of-flight and time-over-threshold of the signal, to allow for time-walk correction. 

The 16 tower control boards are connected to a commercial Zynq-SoC board where the trigger algorithm for an online coincidence selection is implemented. This board is in communication with a computer via a PCIa bus. 
The power consumption of the silicon sensors is 20 mW/cm$ ^{2} $. Since the heat dissipation is distributed all over the towers, a cooling block is implemented in the inter-cell gap to keep the scanner temperature under control. The cooling block is made of laser sintering AlO$ _{3} $ in order to operate within an MRI scanner. The cooling system was designed to keep a very stable and homogenous detector temperature (the uniformity is expected to be at the order of 0.1$ ^{\circ} $C), since the detector temperature is directly linked its mechanical robustness.

The time performance of the detector has been evaluated by measuring the time of flight between two sensors with a beam of 180 GeV pions, available at the CERN SPS beam test facility. The results from the latest prototype, that represents a version of the final chip with a reduced number of channels, showed an average detection efficiency for MIPs of above 99$ \% $ and average time resolution of 130 ps \cite{sensorpaper}.

\section{Scanner simulation}
The first step of the Monte Carlo simulation has been the optimisation of the detection layer structure in order to maximise the $ \gamma $-ray detection efficiency. 
Since the thickness of the scanner ring is constrained by the fact that it has to be inserted in an existing MRI scanner and the thickness of the sensor is fixed  the only free parameters are the thickness and the material of the photon converter. 
Dedicated Monte Carlo studies for the optimisation of the single tower structure were performed with FLUKA \cite{fluka}. This simulation was done with a pencil 511 keV $ \gamma $-ray beam hitting a single tower perpendicularly. Series of simulations for lead, bismuth, uranium and iridium were performed for a single layer thickness from 20 to 100 \textmu{}m, to optimise the detection efficiency. Two materials were chosen for further consideration i.e.: lead and bismuth, both for a single 50 \textmu{}m thick layer. As the results with the two materials were consistent and since the lead is significantly less expensive than the bismuth, a 50 \textmu{}m thick lead converter has been chosen. By dividing the number of photons detected at least in one sensor by the total number of $\gamma$-rays generated, we found a detection efficiency of 27$ \% $. These studies have been crosschecked with GEANT4.\cite{geant4}, which was chosen also for the implementation of the complete scanner simulation. The GEANT4 physics list used is the Em option4 \cite{lowem}.

\subsection{Hit processing and TOF distribution}

\begin{figure}[h!]
\centering
\includegraphics[width=0.55\linewidth]{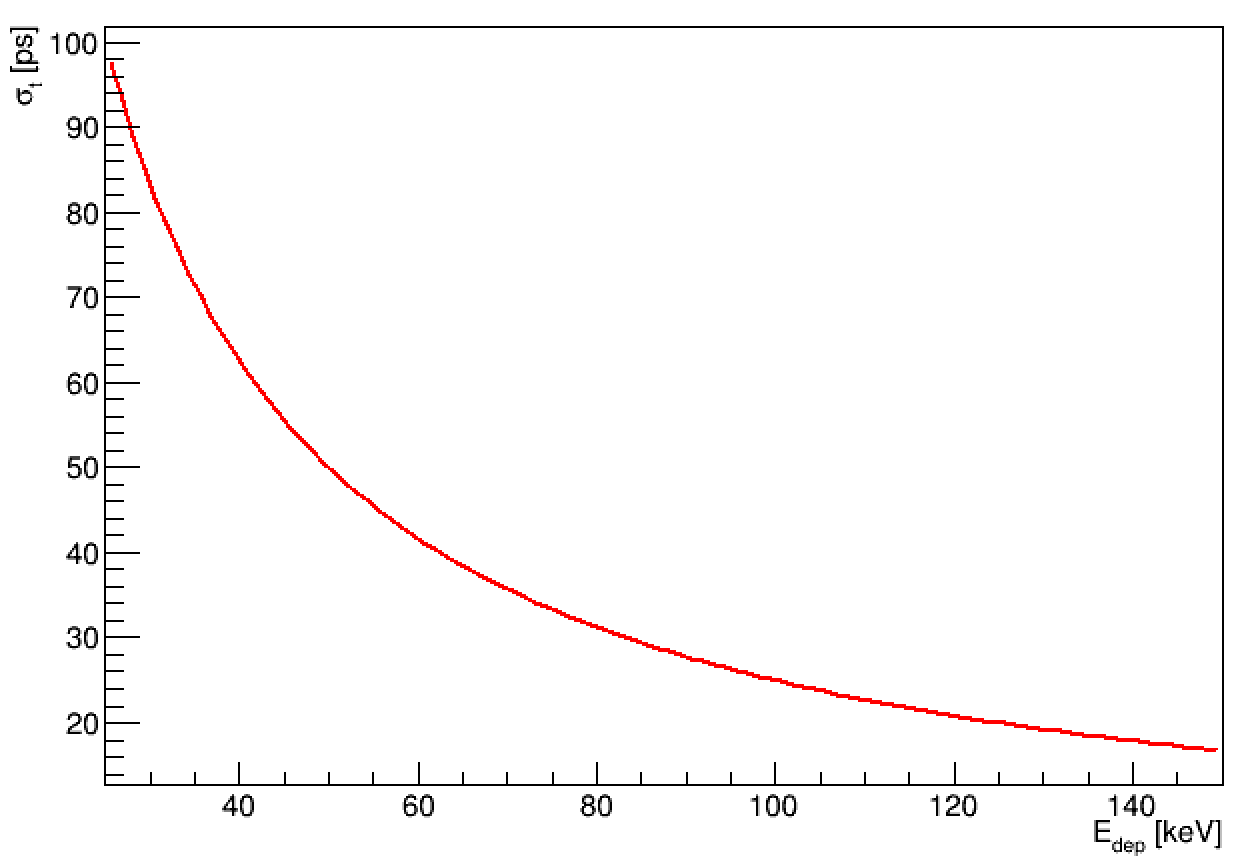}
\caption{Dependence of the time resolution on the energy deposit in the silicon sensor.}
\label{timeresdep}
\end{figure}

The GEANT4 simulation for each disintegration event of positron source provides the hit information in the scanner in terms of position, time and energy deposit in the silicon pixel. This information is processed using a custom software where:

\begin{itemize}

\item The real position of the hit is transformed to the centre of gravity of the pixel where the particle was detected.
\item A gaussian smearing is added to the real time according to the energy dependent time resolution, Fig.~\ref{timeresdep}. \footnote{We assume a  dependence of the time resolution on the energy deposit in the sensor (which is proportional to the charge collected) as $ \sigma_{t}=\sigma_{t}^{0}\frac{E_{dep}^{MIP}}{E_{dep}} $, where $\sigma_{t}^{0}$ is the value measured in \cite{timeres1}.
and $ E_{dep}^{MIP} $ is the most probable energy deposit value for minimum ionising
particles passing through 100 $ \mu m $ of silicon.}
\item The time interval between two disintegrations is distributed as $ e^{-It} $, where I is the intensity of the source. 
\item A DAQ system dead time is introduced to simulate the count losses.

\end{itemize}
\begin{figure}[h!]
\centering
\includegraphics[width=0.55\linewidth]{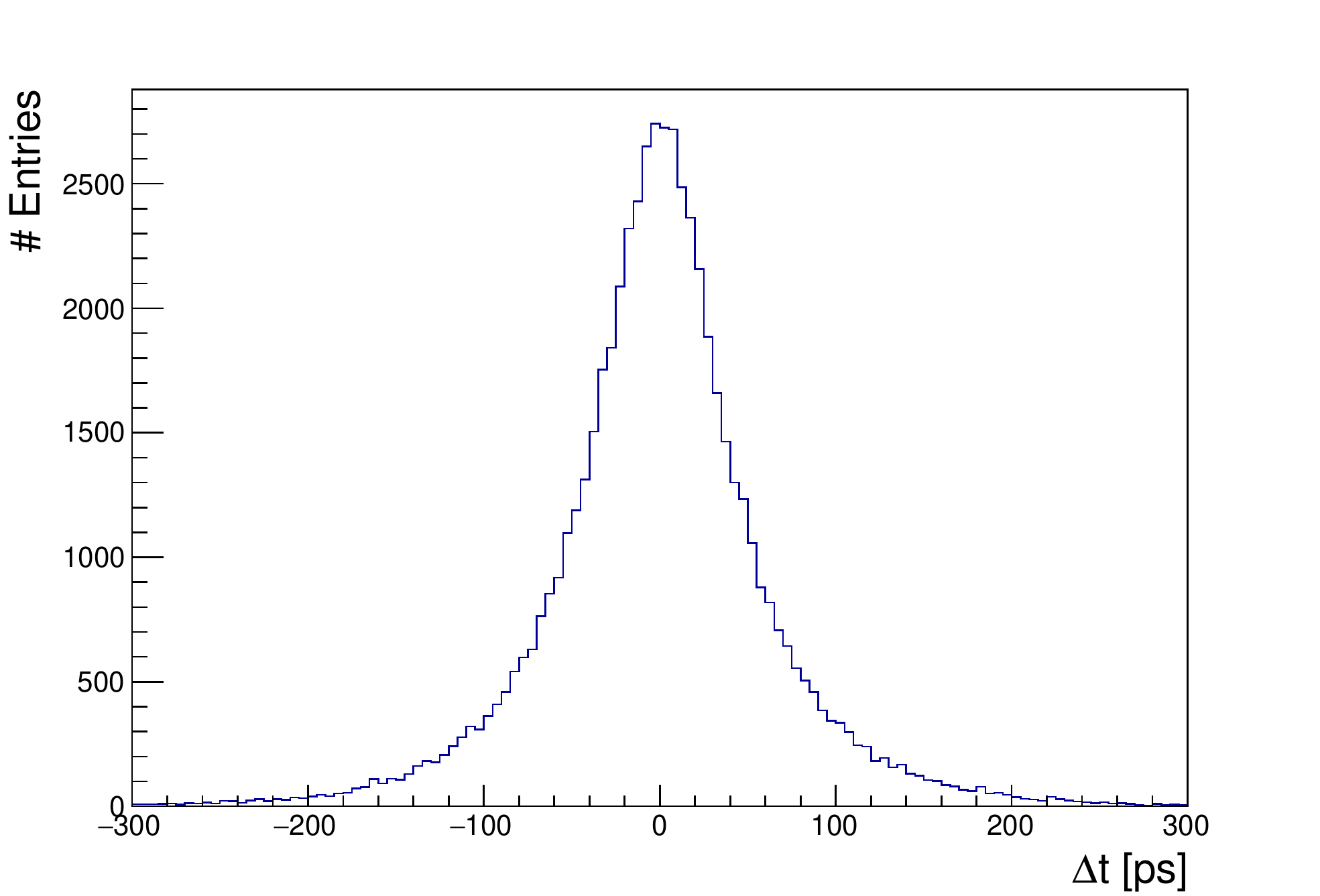}
\caption{Distribution of the time difference between two hits in a coincidence.}
\label{TOF}
\end{figure}

\begin{figure}[h!]
\centering
\includegraphics[width=0.55\linewidth]{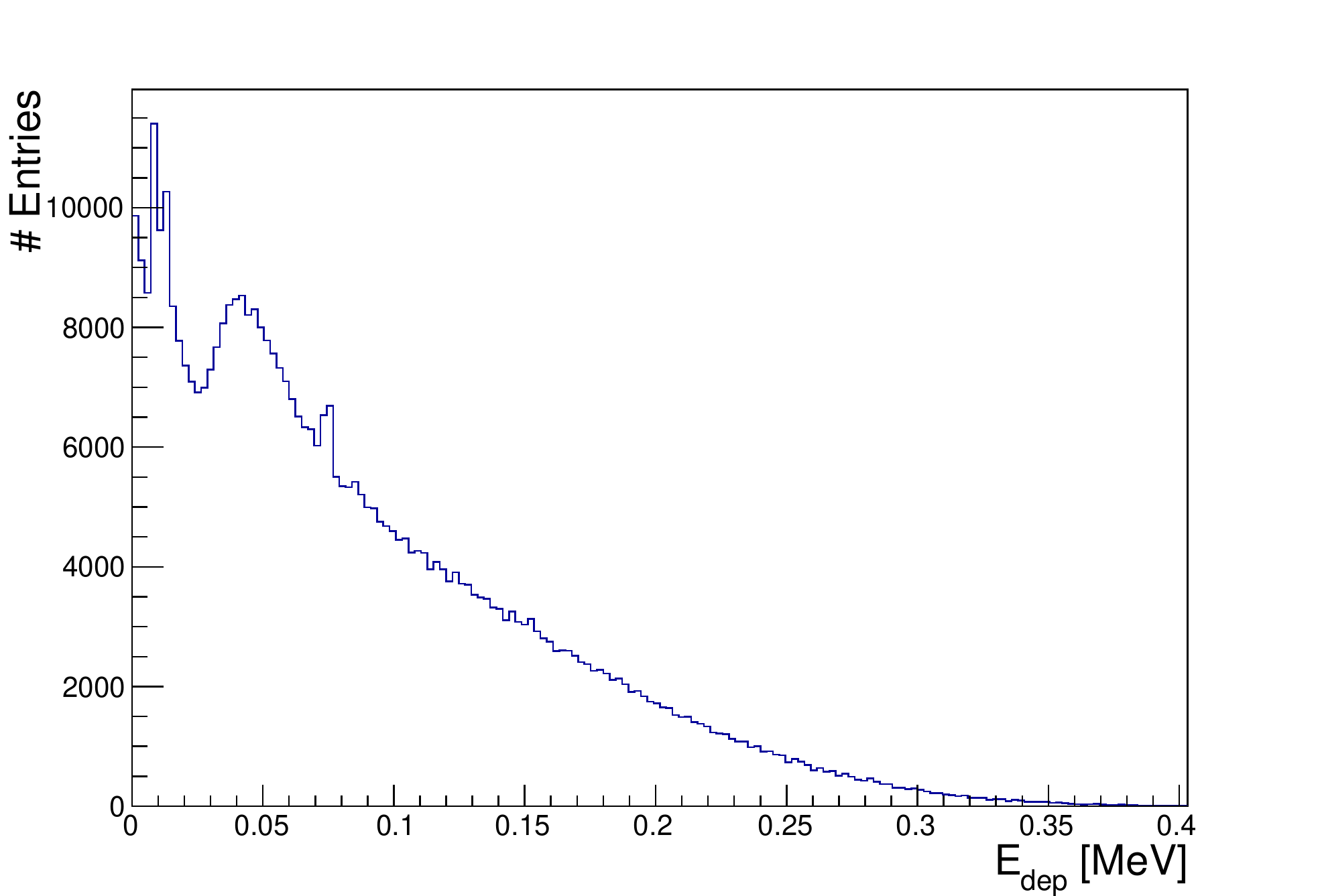}
\caption{Distribution of the energy deposited in the silicon sensor.}
\label{edep}
\end{figure}

To evaluate the expected TOF resolution we simulated a point-like $ ^{18} $F source in an acrylic cube of size 10 mm put in the center of both the axial and the transversal FOV of the scanner.
The distribution of the TOF between two hits in a coincidence is reported in Fig.~\ref{TOF}. The FWHM of the distribution is 80 ps, corresponding to a gaussian standard deviation of 34 ps. Assuming that on average the response of two sensors is the same, the expected time resolution of the single detector is 24 ps, obtained dividing 34 ps by $ \sqrt{2} $. 
The tails are mainly due to the fact that the time resolution of the sensor depends on the energy deposit in the pixel, whose distribution is a continuous spectrum, as shown in Fig.~\ref{edep}. This distribution has a peak at 45 keV, which is mainly due to electrons generated from a photoelectric interaction; the two spikes at 10 keV and 70 keV are due to the production of Auger electrons in silicon; the long tail is mainly due electrons coming from Compton interactions inside the scanner.
The present geometry of the ring does not allow for the measurement of the total energy of a detected $ \gamma $-ray, making it not possible the separation between scattered and true coincidences. Anyway, being the TT-PET scanner designed to operate with small animals, scattered coincidences do not significantly affect the spatial resolution. A higher energy resolution can be achieved by employing thicker sensors and thinner converters and by increasing the number of detection layers. This can be part of future studies for the feasibility of human brain TT-PET scanner.

\subsection{Coincidence rate}

For this study a phantom as described in \cite{nema} sec. 4, a cylinder 50 mm long with a radius of 1.6 mm, has been simulated for a range of source intensity from 5-300 MBq. In the simulation we took into account a 8-state buffer per chip and a dead time of 2 \textmu{}s to transfer the hit from the chip to the tower control. We also considered another dead time of 40 ns, that corresponds to the time needed by the tower control to send one hit to the aggregator. A 500 ps coincidence window was chosen being the time difference for the back $ \gamma $-ray in the scanner of 240 ps, to which 5 times the expected sigma of the time difference distribution. This study has been performed to evaluate also the expected performances of the DAQ system. 
A coincidence event is made by two hits ($hit_{1}$ and $hit_{2}$ ) that satisfies to the following requirements:

\begin{itemize}
\item $|t_{1}-t_{2}|<$ 500 ps
\item The line of response (LOR) intercepts the phantom 
\item The energy deposit in the pixel is larger than 20 keV for both hits.
\end{itemize}

The true Monte Carlo information has been used only to select random and scattered coincidences. With this first cut we include also fake coincidences due to multiple hits in the same layers and due to multiple hits in the same tower. The first case caused by electrons travelling through more than one pixel before stopping while the second case is due to $\gamma$-rays that make Compton scattering in the same tower in which they are absorbed.

\begin{figure}[h]
\centering
\includegraphics[width=0.4\linewidth]{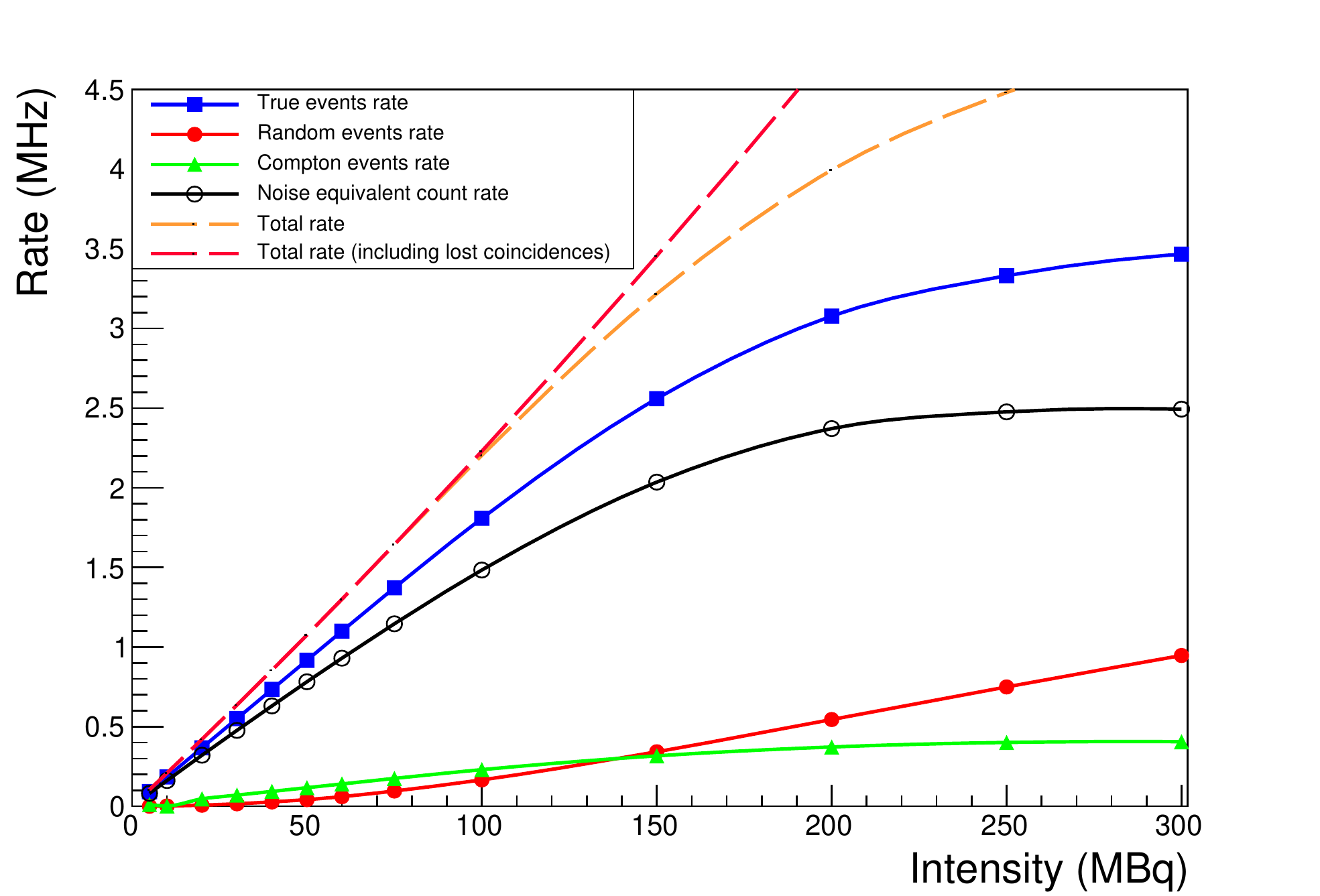}
\includegraphics[width=0.4\linewidth]{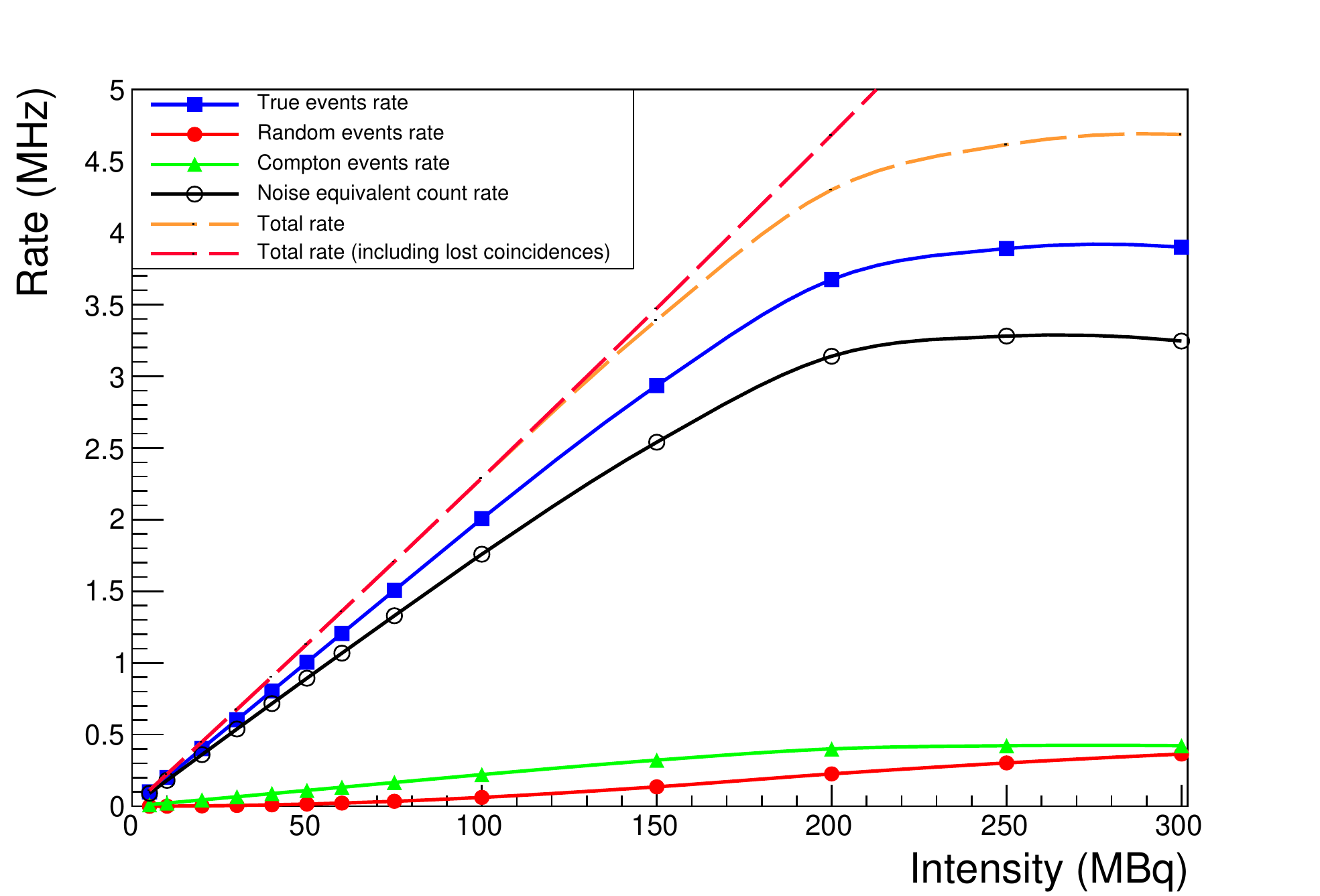}
\caption{(Left) Coincidence rates and NECR as a function of the phantom source activity for a cylindrical source. (Right) Coincidence rates and NECR as a function of the phantom source activity for a spherical source.}
\label{rate}
\end{figure}

To simulate the use case in which we want to scan a limited region of the body of the small animal, we repeated the same study for a spherically distributed source with radius 5 mm, incapsulated in an acrylic cube of side 10 cm located in x=0, y=-5 mm, z=10 mm. The coincidence selection was the same we used for the rod phantom apart that we selected only LORs that intercepted a circle in in the yx plane of radius 8 mm centred in x=0 and y=-5 mm and $|z_{m}-z_{s}|<16 mm$, where $z_{m}$ is the average point of the projection of the line of response along the z axis and $z_{s}$ is the position of the source along the z axis. The results of both simulations are shown in Fig.~\ref{rate}. Comparing the two studies we observe that in the case of the spherical source the random coincidences are reduced by a factor 3. The TT-PET scanner is meant to be operated with a total activity of the source of less than 50 MBq. At this value the count losses are less than 0.1 $\%$ in both cases.

\subsection{Scanner sensitivity}

\begin{figure}[h]
\centering
\includegraphics[width=0.55\linewidth]{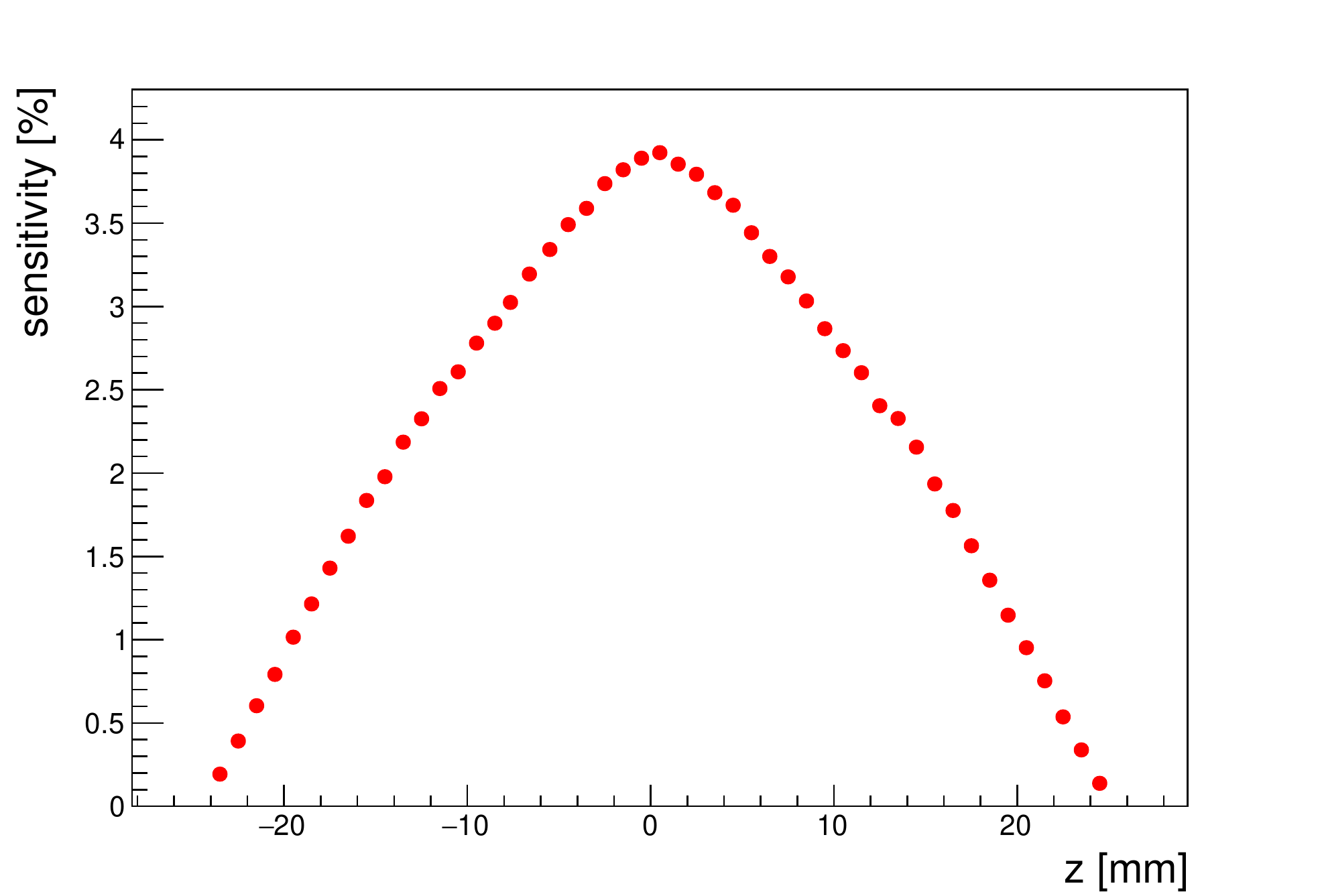}
\caption{Expected sensitivity as a function of the position along the axial FOV}
\label{sensitivity}
\end{figure}

The study a phantom as described in \cite{nema} sec. 5, a spheric source with R=0.3 mm placed at the center of an acrylic cube with a side of 1 cm. For this simulation the source has an intensity of 1 MBq and it has been placed in 49 positions along the z-axis, with steps of 1 mm, in order to cover the entire axial FOV. With this source intensity the random coincidences and the count loss are below the 0.1\%. Fig. \ref{sensitivity} shows the expected sensitivity as a function of the z axis and presents a maximum value at 4 \%.

\section{Reconstruction}

\subsection{Normalisation}
Since the scanner does not have full angular acceptance and the detection efficiency is less than 1, the detection probability of a gamma pair depends on the position in the FOV where it was emitted. The reconstructed images need to be corrected by using a normalisation technique.

\begin{figure}[h]
\centering
\includegraphics[width=0.4\linewidth]{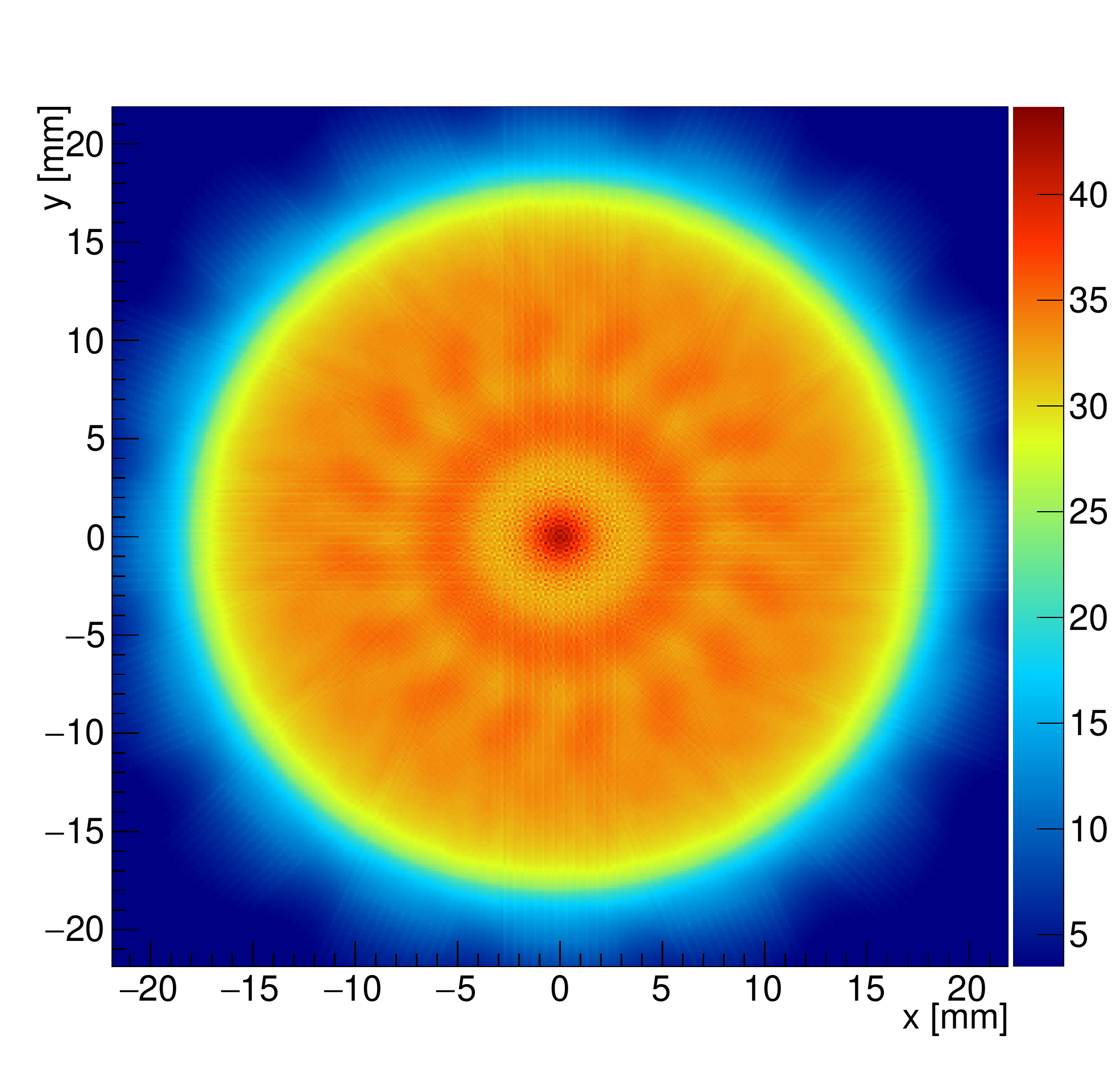}
\includegraphics[width=0.4\linewidth]{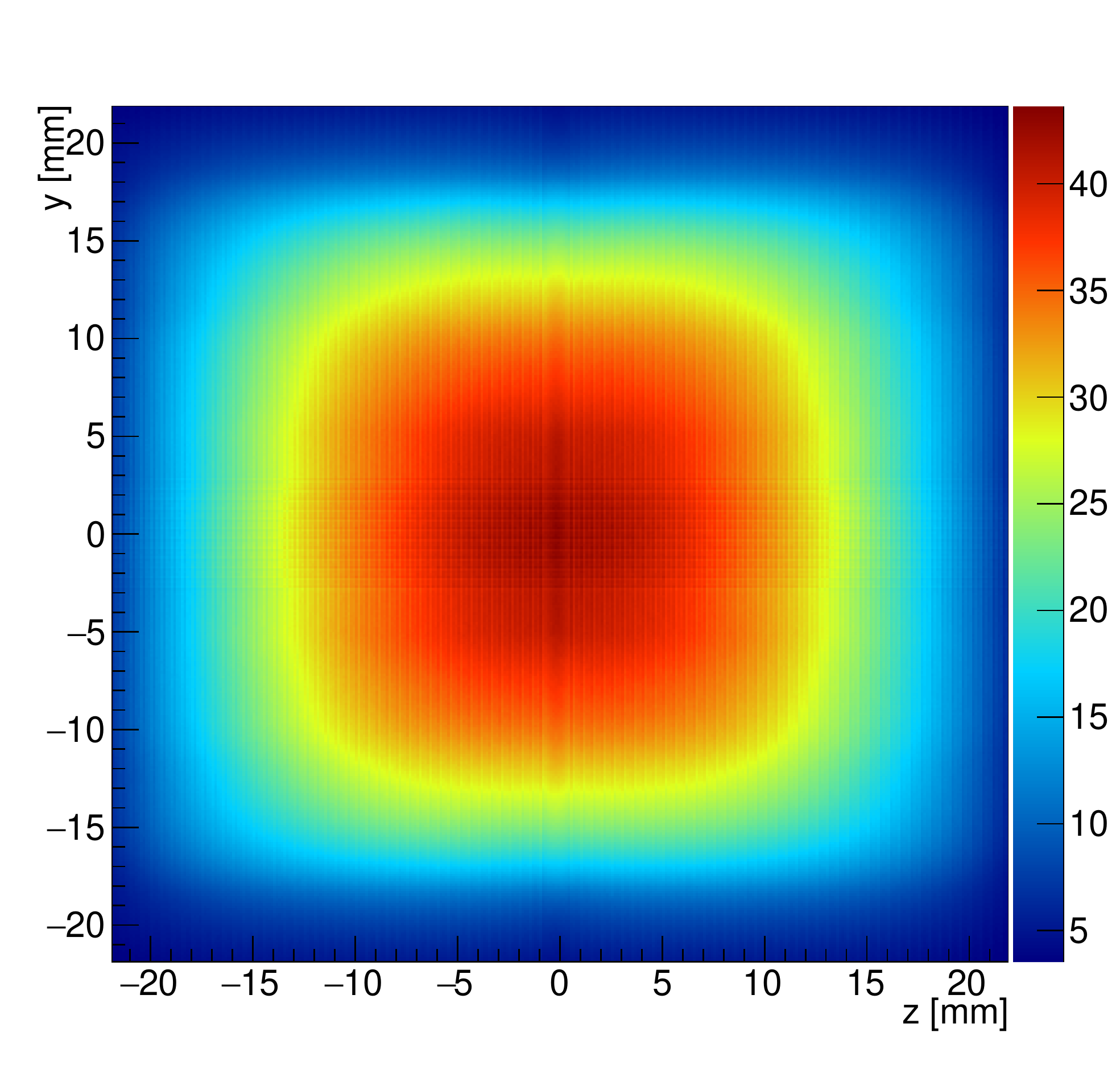}
\caption{From the left Normlaisation map in the y x plane and in the y z plane.}
\label{normalisation}
\end{figure}

\begin{figure}[h]
\centering
\includegraphics[width=0.4\linewidth]{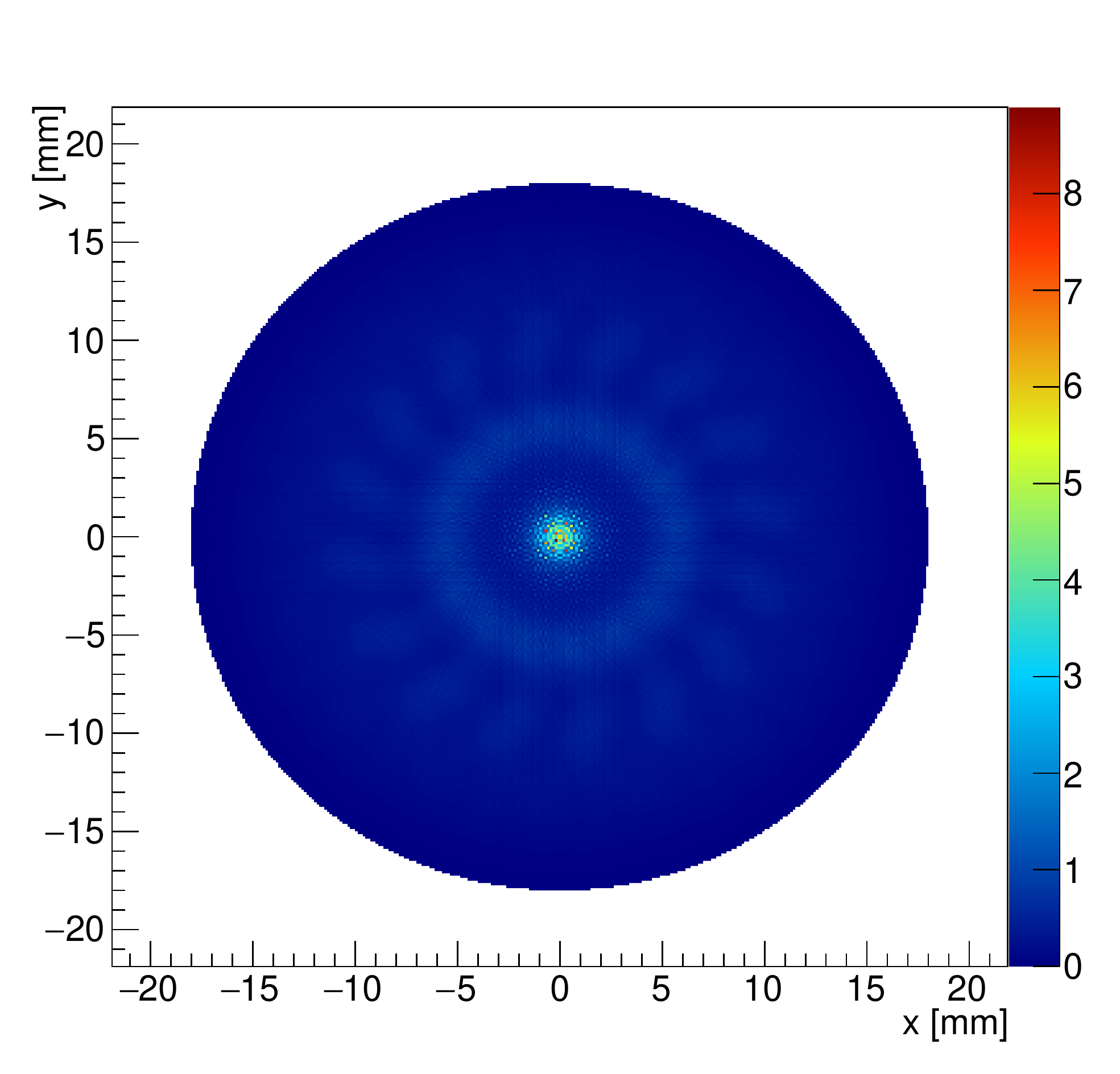}
\includegraphics[width=0.4\linewidth]{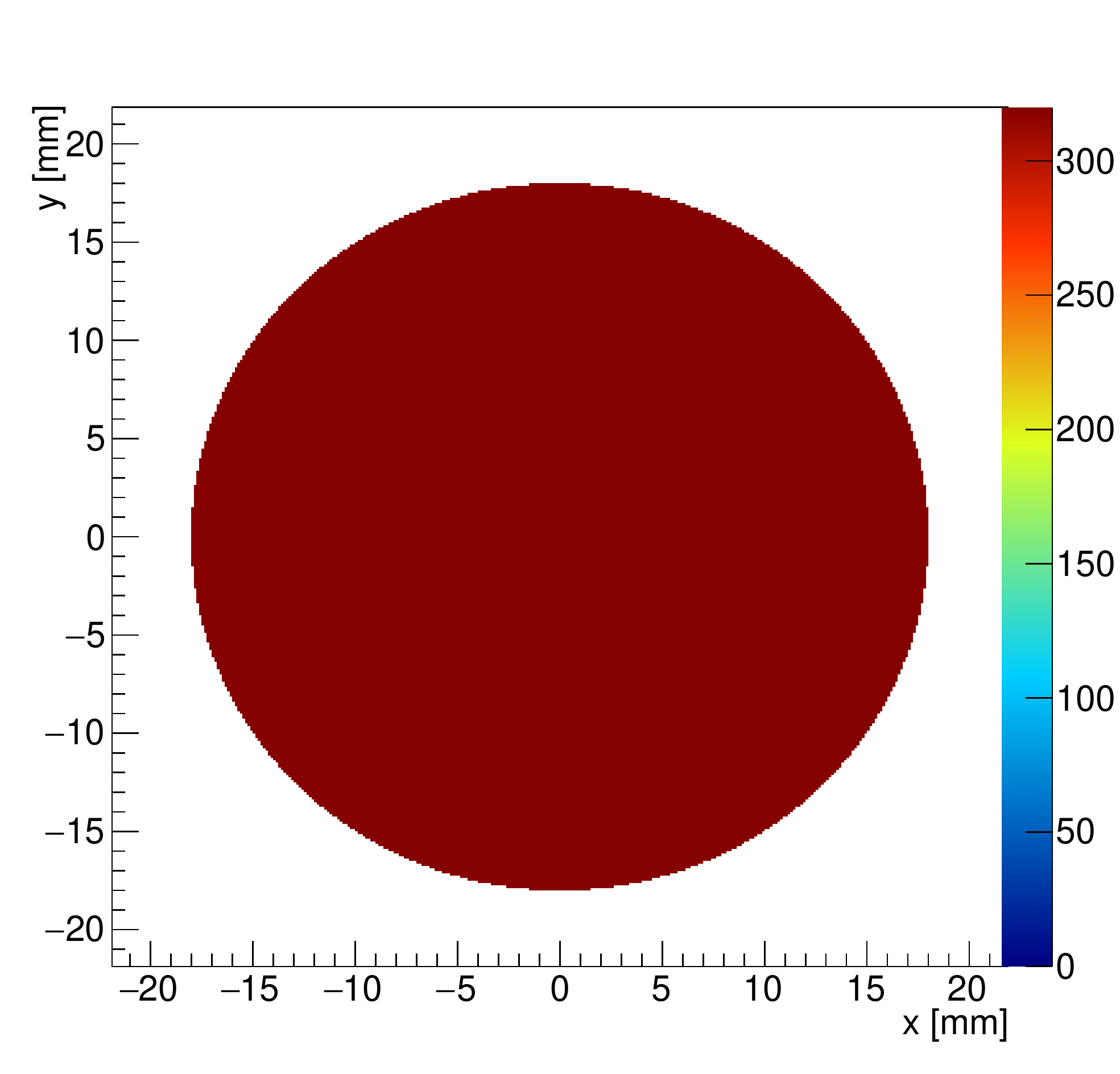}
\includegraphics[width=0.4\linewidth]{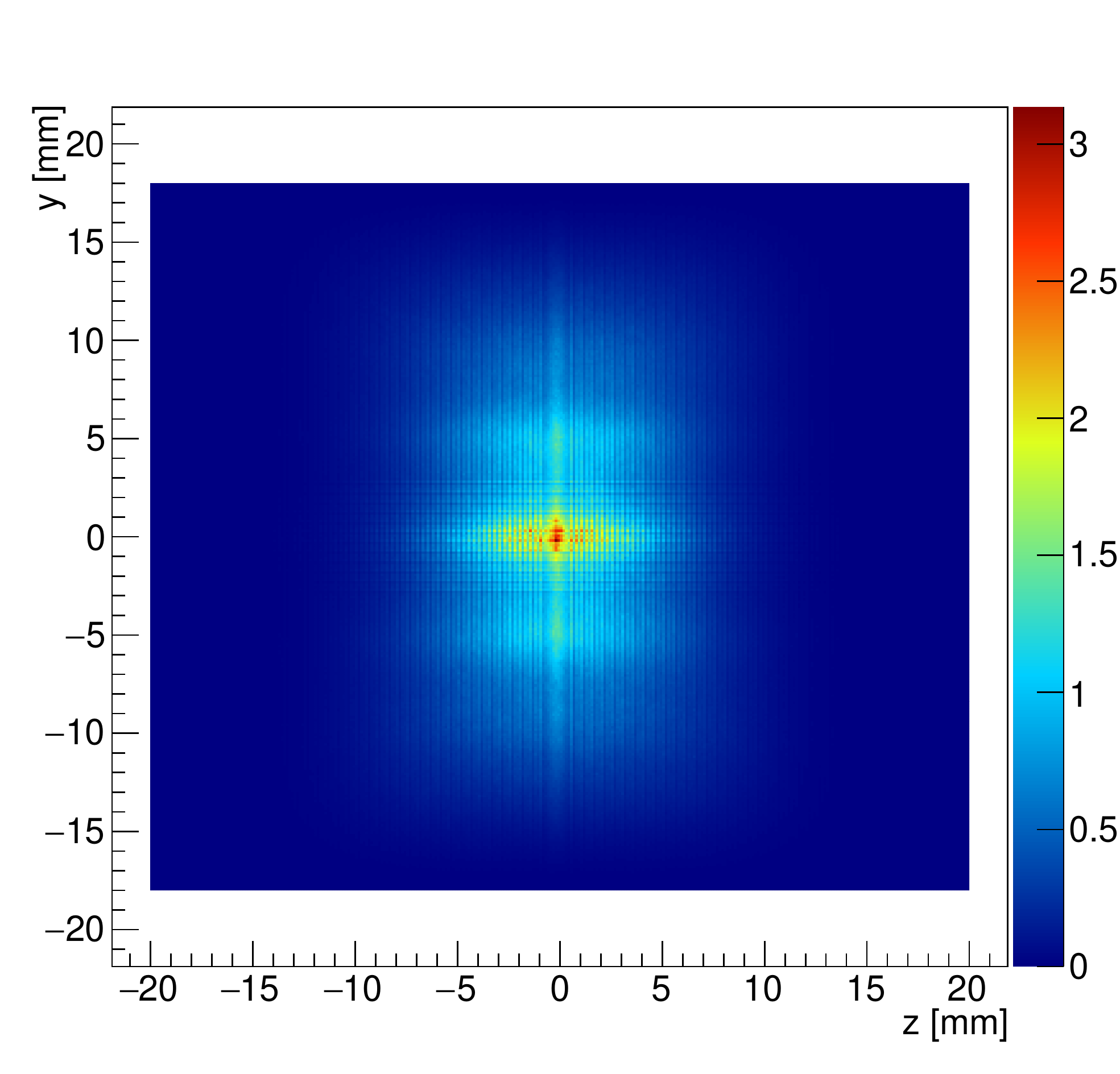}
\includegraphics[width=0.4\linewidth]{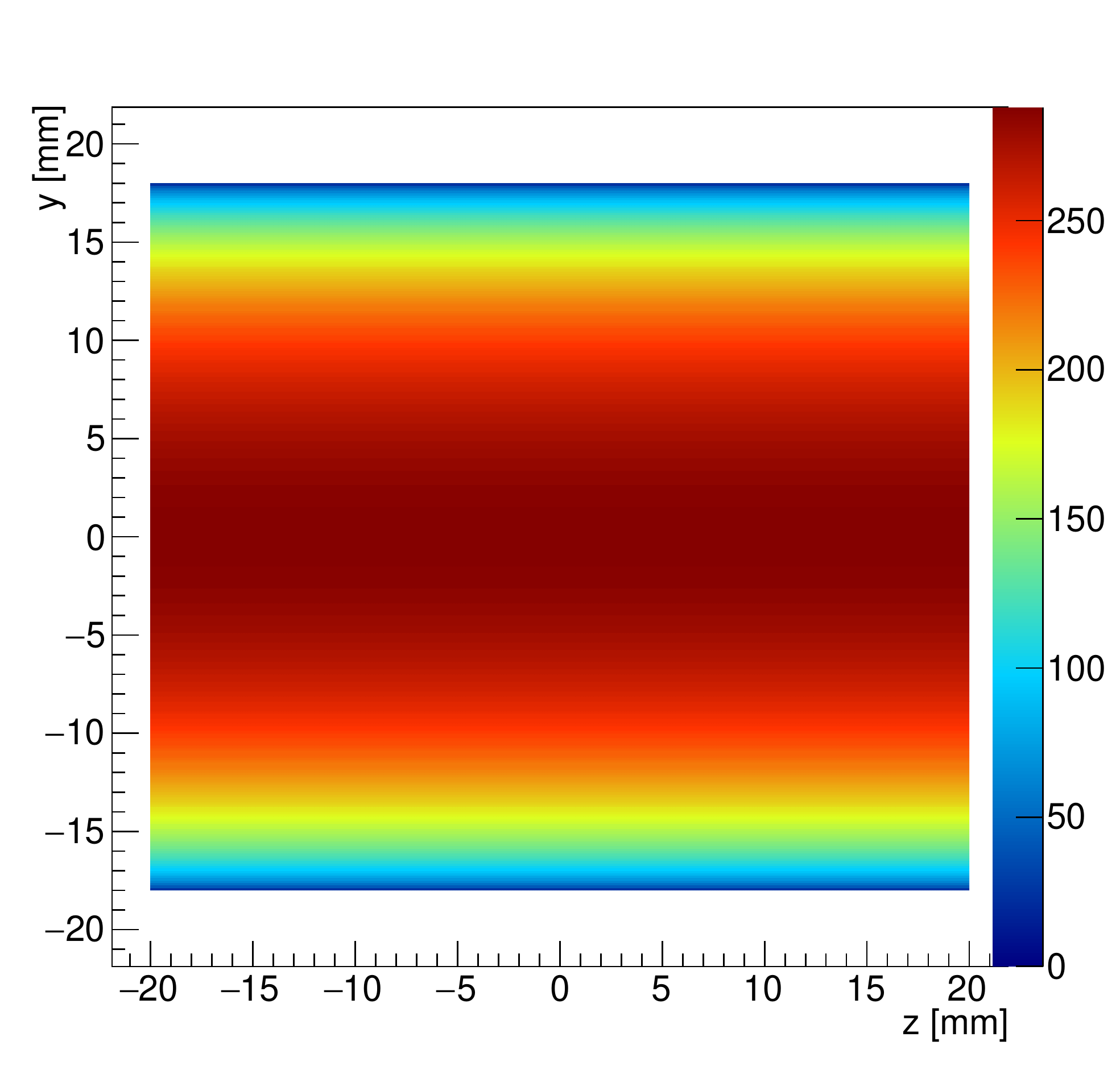}
\caption{(Up) From the left the normalised and not-normalised image in the plane y x. (Down) From the left the normalised and not-normalised image in the plane y z.}
\label{normalised}
\end{figure}

A rod $^{18}$F source with a diameter of 36 mm and length of 40 mm was performed in the simulation and the normalisation factors were calculated as follows;
\begin{align}
Norm_{\mathrm{i}} = I_\mathrm{i}
\end{align}
where $I_\mathrm{i}$ is the intensity of the voxel i-th obtained with an iterative reconstruction of the cylinder. The reconstruction has been performed with the CASToR \cite{CASToR} reconstruction framework based on MELM algorithm. The dimension of the voxel is $0.125 \times 0.125 \times 0.125~mm^{3}$.
Fig.~\ref{normalisation} shows the maps of the normalisation factor obtained with this technique.
Fig.~\ref{normalised} shows the reconstruction of the image of the uniform cylinder in a region of interest R<18 mm and |z|<20 mm, that corresponds to the volume where the source is distributed, before and after the normalisation.

\subsection{Spatial resolution}

\begin{table}[h]
\centering
\begin{tabular}{lcccc}
\hline 
\textbf{x position {[}mm{]}} & \textbf{0} & \textbf{5} & \textbf{10}& \textbf{15} \tabularnewline
\hline 
\textbf{FWHM radial [mm]} & 0.59 &  0.57 & 0.56 & 0.52 \\
\textbf{FWHM tangential [mm]} & 0.60 & 0.60 &  0.67 & 0.71 \\ 
\textbf{FWHM axial [mm]} & 0.50 & 0.49 & 0.5  & 0.51    \\
\textbf{FWTM radial [mm]} & 1.8 & 1.6 & 1.5 & 1.4   \\
\textbf{FWTM tangential [mm]} & 1.8 & 1.7 & 1.9  & 2.0  \\  
\textbf{FWTM axial [mm]} & 1.2 & 1.1 & 1.1  & 1.1   \\
\hline 

\end{tabular}

\caption{\label{tab:tab1}Spatial resolution values along the transverse FOV at the center of the axial FOV.}
\end{table}

\begin{table}[h]
\centering
\begin{tabular}{lcccc}
\hline 
\textbf{x position {[}mm{]}} & \textbf{0} & \textbf{5}& \textbf{10}& \textbf{15} \tabularnewline
\hline 
\textbf{FWHM radial [mm]} & 0.65 & 0.61 & 0.60  & 0.56  \\
\textbf{FWHM tangential [mm]} & 0.64 & 0.65 & 0.65 & 0.7\\
\textbf{FWHM axial [mm]} & 0.45 & 0.45 & 0.45 & 0.45 \\
\textbf{FWTM radial [mm]}  & 2.0 & 1.8 & 1.7 & 1.6 \\
\textbf{FWTM tangential [mm]} & 2.0 & 1.9 & 1.9 & 2.0 \\
\textbf{FWTM axial [mm]}  & 0.94 & 0.96 & 1.0 & 1.0 \\
\hline 

\end{tabular}

\caption{\label{tab:tab2}Spatial resolution values along the transverse FOV at 1/4 of the axial FOV (z=12.5 mm).} 
\end{table}

The same phantom used for the sensitivity study has been simulated in the position 0, 5, 10, 15 at the center of the axial FOV and 0, 5, 10, 15 at z=12.5 mm as described in \cite{nema} sec. 3. For each simulation more than 100000 coincidences have been selected with count losses and random coincidences less than 0.1$\%$. The point spread function (PSF) has been reconstructed using the CASToR \cite{CASToR} reconstruction framework. After 2 iterations we found a result compatible with 2DFBP. Both the FWHM and the FWTM has been evaluated for each point spread function in the axial, tangential and radial component. The results are shown in Tab.~\ref{tab:tab1} and Tab.~\ref{tab:tab2}
\begin{figure}[h]
\centering
\includegraphics[width=0.5\linewidth]{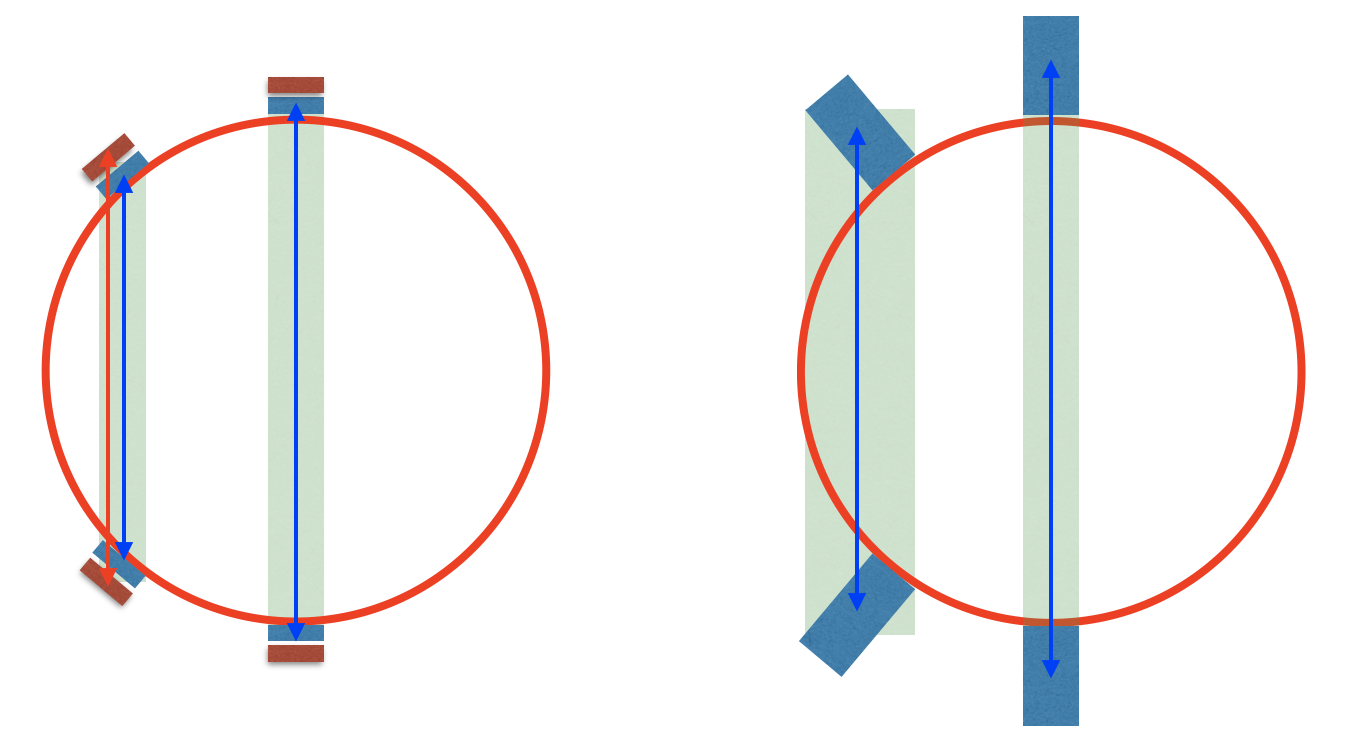}
\caption{Schematisation of the precise DOI measurement.}
\label{degradation}
\end{figure}

Apart from variations of less than 10\%, we notice how the resolution is constant in the different positions along the transversal and the axial FOV and they it is below 0.75 mm. As we can see from Fig~\ref{degradation}, the usual degradation of the spatial resolution at high values along the radial direction \cite{petcomparison} is recovered thanks to the multi layer structure.

\subsection{Derenzo phantom reconstruction}

\begin{figure}[h]
\centering
\includegraphics[width=0.4\linewidth]{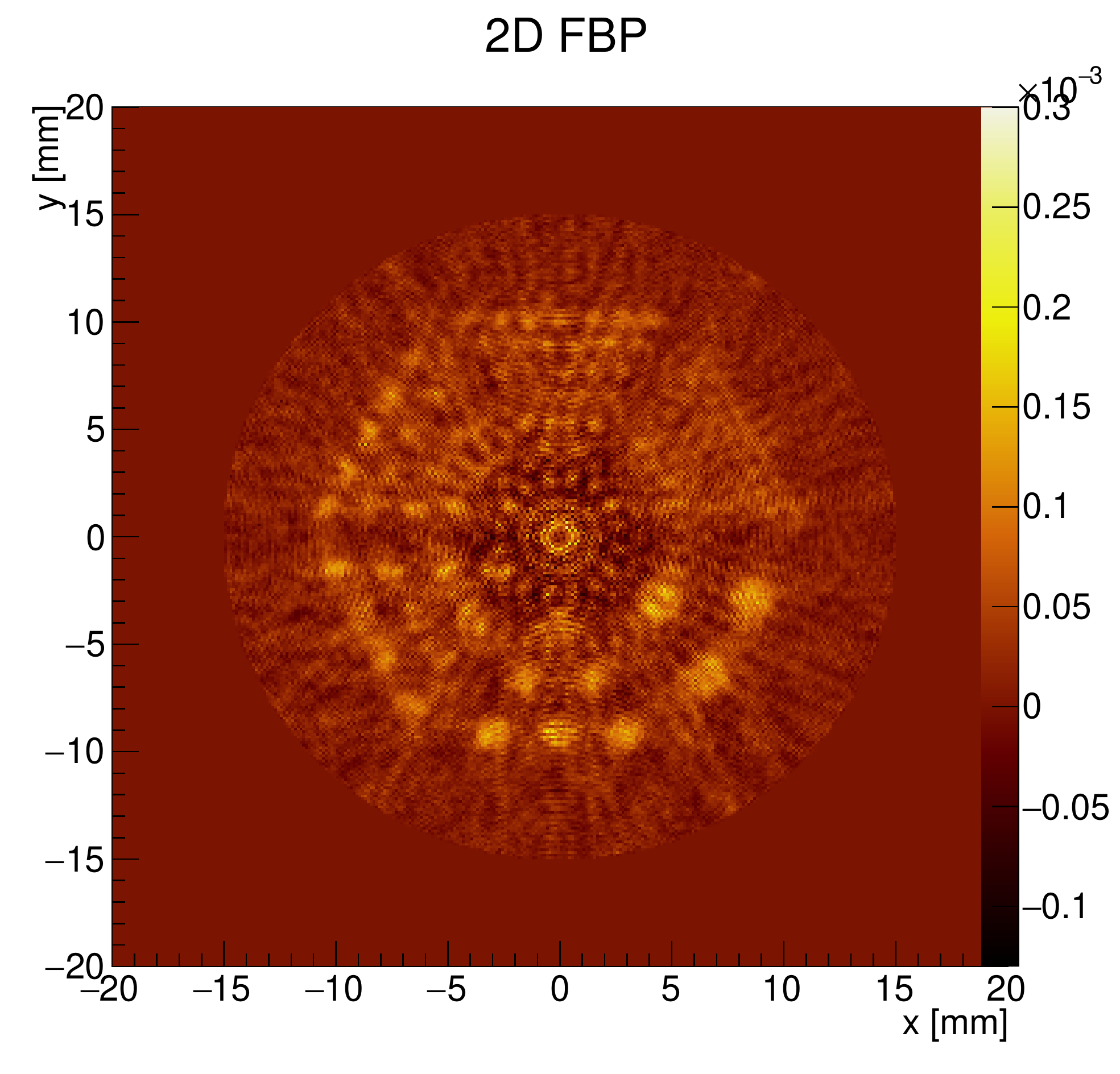}
\includegraphics[width=0.4\linewidth]{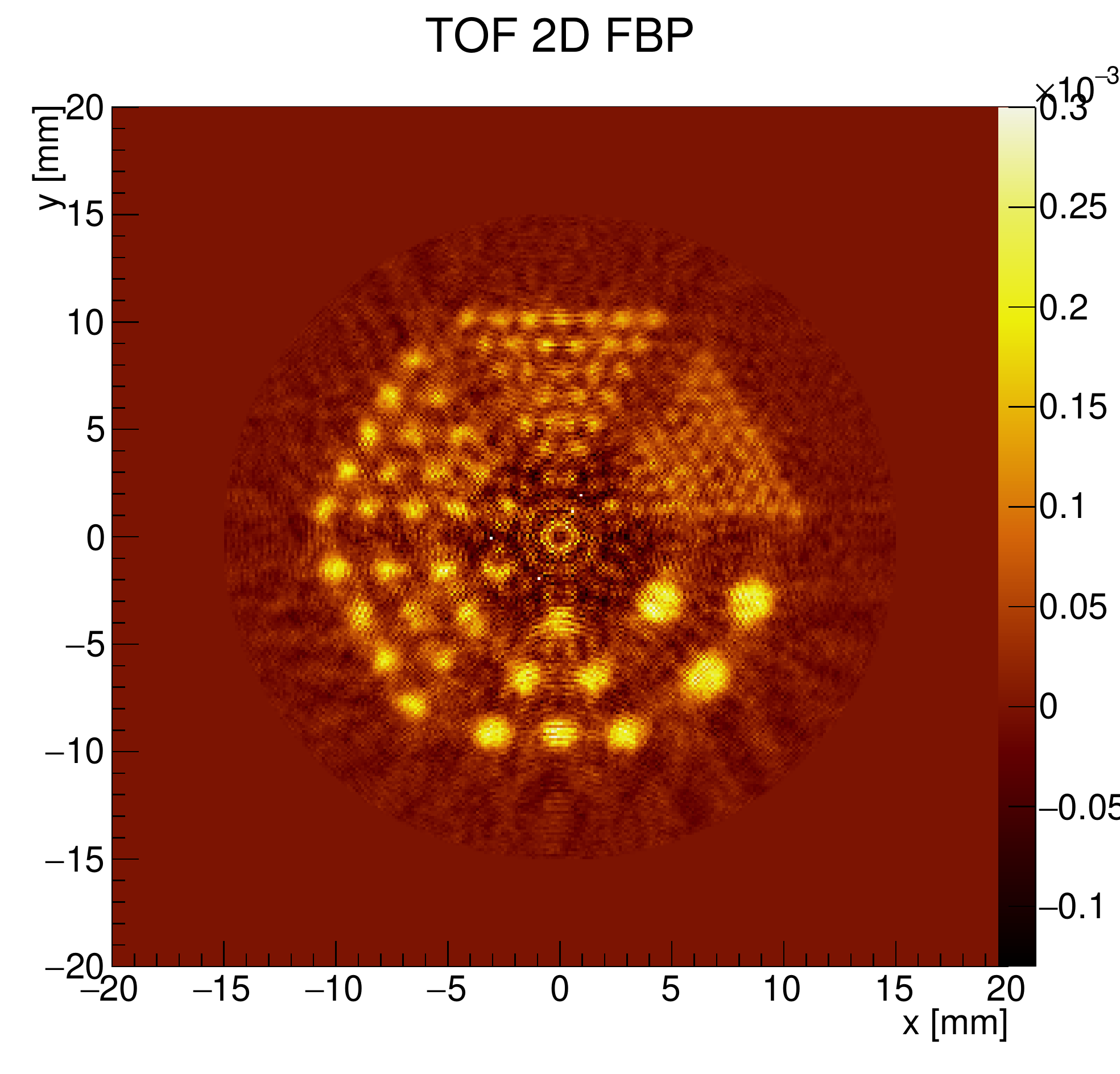}
\caption{(left) The Derenzo phantom reconstructed without TOF information and (right) with TOF information.}
\label{derenzo}
\end{figure}

We simulated a Derenzo phantom with 40 mm long rods of different diameter: 0.5 mm, 0.7 mm, 1 mm, 1.2 mm, 1.5 mm, 2 mm. The distance between the rods with the same diameter is two times the diameter itself. The total intensity of the phantom is 50 MBq. 
The reconstruction was done using a custom software based on filtered back projection (FBP), rebinning the z axis in 20 slices of 2 mm according to the single slice rebinning method \cite{singleslice}. Fig.~\ref{derenzo} shows how the TOF information significantly improves the signal to noise ratio. 

\section{Discussion}
A time resolution of 100 ps has been measured for minimum ionising particles using the silicon pixel detector developed for the TT-PET scanner. Starting from this value, a detailed Monte Carlo simulation showed that a value of 24 ps is expected for 511 keV  $\gamma$-ray. The final TT-PET sensor has a monolithic architecture, configuration that has the main advantage of covering a large active surface with a much lower cost with respect to a hybrid system, where the sensor bonded to the front-end ASIC. The high granularity of the detection modules and the organisation of the detection layers in super-modules turns out in a smooth saturation of the read-out system, which is able to acquire coincidences with a count loss lower than 0.1$\%$ up to 75 MBq activity.
Dedicated simulations showed that thanks its high 3D granularity the scanner has an expected average spatial resolution of 650 \textmu{}m FWHM, with a uniform response in whole FOV.

\section{Conclusion}

The TT-PET project introduces the use of fast silicon monolithic detector in the TOF-PET scanner technology. We are currently developing a small animal scanner, meant to be inserted in an existing MRI machine. The excellent performance in terms of rate capability and space resolution depend on the development of the sensor ASIC. The first demonstrator of the TT-PET chip was successfully tested showing a time resolution of 130 ps for minimum ionising particles. We are currently testing the same chip with a $^{22}$Na source for measuring the time performance for 511 keV photons. Concerning the simulation and reconstruction activity, we are now working with the CASToR community to adapt the reconstruction software to the high 3D granularity and time-of-flight of the scanner.

\section{Aknoledgements}
This research is supported by the Swiss National Science Foundation grant CRSII2-160808. We would like to thank our colleagues from the University of Brest (FR) for the invaluable help and support.


\end{document}